\documentclass{cs20proc}

\usepackage{kantlipsum}

\editors{S. J. Wolk}
\publisher{Zenodo}
\conference{The 20th Cambridge Workshop on Cool Stars, Stellar Systems, and the Sun}
\conferencedate{2018}

\title{The Dissolution of Clusters: What Can We Learn from Nearby, UV-bright, Overluminous Field Stars?}
\author{Joel Kastner$^{1}$, Matthieu Chalifour$^{1,2}$, Alex Binks$^{3}$, David Rodriguez$^{4}$, Kristina Punzi$^{5}$, G. G. Sacco$^{6}$}

\affiliation{	$^{1}$Center for Imaging Science and Laboratory for Multiwavelength Astrophysics, Rochester
                         Institute of Technology, Rochester, NY, USA \\ 
$^{2}$Department of Physics \& Astronomy, Swarthmore
  College,  Swarthmore, PA, USA\\
                       $^3$Astrophysics Group, School of Chemistry and Physics, Keele University, UK \\
                     $^4$Space Telescope Science Institute, Baltimore, MD, USA \\
                   $^5$Astronomy Department, Wellesley College, Wellesley, MA, USA\\
                 $^6$Arcetri Observatory, Florence, Italy}

\shorttitle{UV-bright, Overluminous Field Stars}
\shortauthors{Chalifour, Binks, Kastner, et al.}

\abs{The past two decades have seen dramatic progress in our knowledge
  of the population of young stars of age $<$200 Myr that lie within
  150 pc of the Sun. These nearby, young stars, most of which are
  found in loose, comoving groups, provide the opportunity to explore
  (among many other things) the dissolution of stellar clusters and
  their diffusion into the field star population. In the age of {\it Gaia},
  this potential can now be fully exploited. We have identified, and
  are now investigating, a sample of nearly 400 {\it Galex} UV-selected
  late-type (K and early-M) field stars with {\it Gaia}-based distances $<$120
  pc and isochronal ages $\le$80 Myr (even if binaries). Only a
  small percentage ($<$10\%) of stars among this
  (kinematically unbiased) sample can be confidently associated with
  established nearby, young moving groups (NYMGs). The majority display anomalous
  kinematics, relative to the known NYMGs. These stars may hence
  represent a previously unrecognized population of young stars that
  has recently mixed into the older field star population. We
  discuss the implications and caveats of such a hypothesis---including the intriguing fact that, in addition to their
  non-young-star-like kinematics, the majority of the UV-selected,
  isochronally young field stars within 50 pc appear surprisingly
  X-ray faint.} 

\begin{document}

\maketitle

\section{Introduction}

The identification and study of stars of age $<$200 Myr 
within $\sim$100\,pc of the Sun
provides crucial insight into pre-main sequence (pre-MS) stellar
evolution and the formative stages of planetary systems \citep{Kastner2016IAUSed}. Such young,
nearby stars provide excellent samples for direct-imaging campaigns to
observe exoplanets and circumstellar discs
\citep[e.g.,][]{2004a_Kalas, 2008a_Lagrange, 2015a_Bowler,
  2015a_MacGregor, 2015a_Chauvin}, act as direct observational
test-beds for early stellar evolution \citep[e.g.,][]{2004a_Zuckerman,2015a_Bell}, and provide key
evidence for the physical origins of young stars in the Solar
neighbourhood \citep[e.g.,][]{WrightMamajek2018}. 

The majority of these nearby, young stars can be placed in
kinematically coherent ensembles known as nearby young moving groups
(NYMGs). To date, at least a dozen, and perhaps as many as two dozen,
NYMGs have been identified \citep{Mamajek2016,2018a_Gagne}. Since NYMGs
are approximately coeval \citep[age spreads generally
$< 5$\,Myr;][]{2015a_Bell}, any age determination methods for a
star in a NYMG can reasonably be applied to any other star in the
group; furthermore, age determinations from diverse methods can create
a tight age constraint for the NYMG \citep[e.g.,][]{MamajekBell2014}. Recent work suggests that NYMGs
share similar chemical abundances
\citep{2013a_DeSilva,2013a_Barenfeld}, which provides
evidence for their common origins and hints at the compositions of the
molecular clouds from which they were born.

Over the past two decades, the identification of candidate nearby,
young, late-type stars and (hence) NYMG members among the field-star
population has proceeded via some combination of their
signature luminous chromospheric (UV) and coronal (X-ray) emission, which result from strong
surface magnetic fields \citep[e.g.,][and
references therein]{1997a_Kastner,2013a_Rodriguez}, and their common
Galactic ($UVW$) space motions \citep[e.g.,][]{2004a_Zuckerman,2008a_Torres,2013a_Malo}. Follow-up
spectroscopy then further constrains stellar ages via determinations of Li
absorption line strengths, rotation rates, and optical activity
indicators (such as H$\alpha$ and Na I emission lines), so as to assess the viability of candidate NYMG stars or of
proposed new NYMGs \citep[see discussions in][]{2004a_Zuckerman, 2015a_Binks}.

The fact that so many field stars, even within the nearest 100\,pc, were missing parallaxes, precise proper motions, and/or radial velocity measurements presented a major difficulty for previous searches for NYMG candidates and tests of their membership status \citep[e.g.,][]{2014a_Malo}. With the sudden availability of such data, in the form of the first two data releases from the {\it Gaia} space astrometry mission \citep[Data Releases 1 and 2, hereafter DR1 and DR2;][]{2016a_Gaia_Collaboration,2018a_Gaia_Collaboration}, the study of nearby, young stars and NYMGs can now make major strides. This potential motivated the recent study, described in \citet[][]{2017a_Kastner}, in which we evaluated the distances and ages of all 19 nearby young star candidates from the sample of $\sim$2000 stars compiled by the {\it Galex} (UV) Nearby Young Star Survey \citep[GALNYSS;][]{2013a_Rodriguez} that were included in {\it Gaia} DR1. The youth of the majority of these 19 mid-K to early-M stars was confirmed by their positions, relative to both the loci of main sequence stars and theoretical isochrones, in {\it Gaia} color-magnitude and color-color diagrams. Surprisingly few of the GALNYSS stars included in {\it Gaia} DR1 have kinematics consistent with membership in known NYMGs, however \citep{2017a_Kastner}.

In the present work, we further investigate the ability of {\it Gaia} data to identify nearby, young stars and to assess their NYMG memberships --- or lack thereof. Guided by the \citet{2017a_Kastner} study, we have used {\it Gaia} DR1 to select a sample consisting of a few hundred bright ($7 < V < 12.5$) stars with {\it Galex} UV counterparts that are isochronally young (ages $\sim$80 Myr). We then used {\it Gaia} DR2 data to assess these stars' kinematics. For subsamples of these {\it Gaia}/{\it Galex}-selected nearby young star candidates, additional archival data  (e.g., X-ray emission and Li absorption) have been compiled with which we can assess diagnostics of youth. Here, we discuss the main characteristics of this isochronally and UV-selected sample of candidate nearby, young stars. The details, including tables listing ({\it Gaia}/{\it Galex}/2MASS) astrometric, photometric, and kinematic data for the full sample\footnote{Preliminary versions of these tables can be
  obtained, on request, from the authors.}, will be included in a
forthcoming paper (Chalifour et al.\ 2019, in prep.).

\section{Selecting Candidate Nearby, Young Stars from {\it Galex} and
  {\it Gaia} Data}

Expanding on \citet{2017a_Kastner}, our initial selection of stars for the present study was based on crossmatching {\it Gaia} DR1 catalog entries\footnote{The present study was initiated before the release of {\it Gaia} DR2, and so relies entirely on DR1 for sample selection. However, for the analysis described in \S 3, we use DR2 astrometric and photometric data. This substitution is justified by the fact that, for our final sample of 376 objects, the absolute difference between DR1 and DR2  parallaxes is less than twice the combined error bar from DR1 and DR2 for 95\% of the sample stars, and this difference is never larger than 5.0\,pc.} with the {\it Galex} All-sky Imaging Survey (AIS) point source catalog, but without the additional proper motion constraints used by \citet[][]{2013a_Rodriguez} to assemble the GALNYSS catalog. We adopted a cross-matching radius of 3$''$ to associate {\it Gaia} DR1 entries with NUV photometry from the {\it Galex} AIS and, subsequently, near-IR ($JHK_{\rm s}$ ) and mid-IR ($W1$--$W4$) photometry from 2MASS and WISE, respectively, using the Vizier crossmatch service\footnote{http://cdsxmatch.u-strasbg.fr/xmatch}. This cross-matching exercise generated a catalog with 715,773 objects. 

\begin{figure}[!h]
\centering
\includegraphics[width=0.48\textwidth,angle=0]{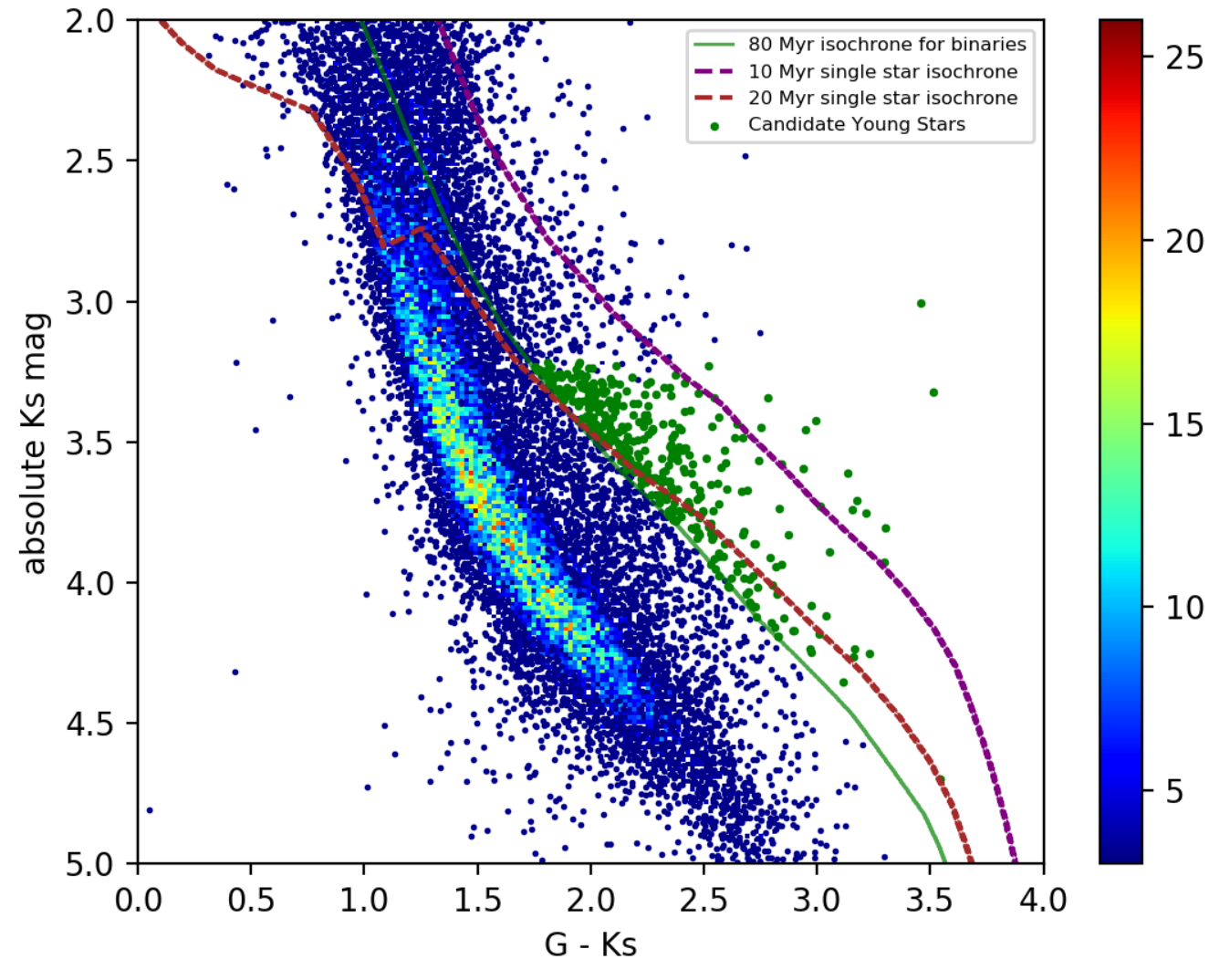}
\includegraphics[width=0.48\textwidth,angle=0]{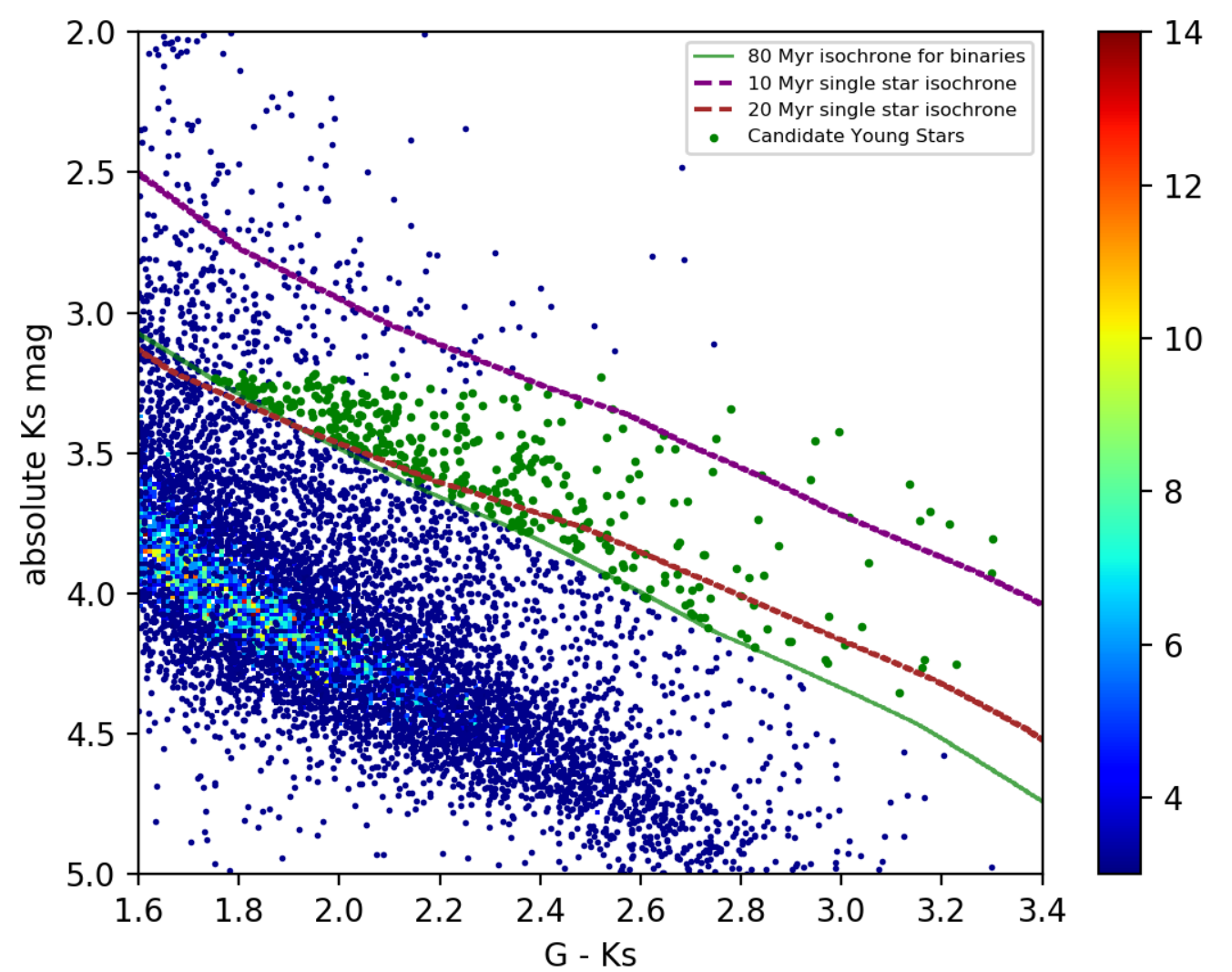}
\caption{Top: $K_{\rm s}$ vs.\
  $G-K_{\rm s}$ color-magnitude diagram (density plot) of {\it Galex} UV-selected DR1 stars, with positions of our 376 young
  star candidates indicated as green
  circles. Theoretical pre-MS isochrones from T11 for ages
  of 10 and 20 Myr as well as 80 Myr ``binary stars'' are overlaid;
  i.e., the 80 Myr isochrone has been adjusted upwards by 0.75 mag, to
  simulate the positions of equal-components binaries of that
  age. Bottom: zoomed-in view of the same CMD, centered on the positions of the candidates.
}
\label{fig:GaiaColMag}
\end{figure}

We then selected stars within 125 pc (i.e., parallaxes $\pi \geq 8\,{\rm mas}$) that lie significantly above the main sequence (MS) according the models of \citet[][hereafter T11]{2011a_Tognelli}. Specifically, the T11 isochrones were used to select the subset of stars that appear younger than 80\,Myr --- even if equal-components binaries \citep[see, e.g., ][]{2017a_Kastner} --- in a $M_{K_{\rm s}}$ vs $G-K_{\rm s}$ color-magnitude diagram (Fig.~\ref{fig:GaiaColMag}). The evolutionary models of \citet{2015a_Baraffe} and those of T11 agree to within a few tenths of a magnitude at 80\,Myr for K and early-M type stars, with the T11 models consistently predicting older ages for low-mass stars \citep{2017a_Kastner}. For this reason, the T11 models are a more conservative choice for selecting stars younger than 80\,Myr.
To further limit the sample size, we then selected only stars lying below (less luminous than) and redward of the T11 $1.0$ $M_{\odot}$ evolutionary track. No lower limit on mass was imposed, although the use of the Tycho catalogue to construct the DR1 catalog imposes a magnitude limit of $V \sim 9$, which should result in a sample dominated by young K and early M dwarfs (\S 3.1). 

The foregoing selection criteria resulted in the sample of 376 stars highlighted in Fig.~\ref{fig:GaiaColMag}. These candidate young stars are more or less uniformly distributed across the sky, with the exception of the {\it Galex} Galactic plane avoidance zone. The candidate sample includes a small number of previously identified NYMG members (\S 3.3.2) --- including TW Hya, the namesake of the $\sim$10 Myr-old association whose identification spawned the wider search for NYMGs and their members \citep{1997a_Kastner,2004a_Zuckerman,2008a_Torres}.

\section{Properties of the Candidate Stars}

\subsection{Spectral types} 

We assigned spectral types to the sample stars either on the basis of the stars' entries in the SIMBAD astronomical database \citep[152 stars;][]{2000a_Wenger} or via optical/near-IR colors. For the latter stars, we used the average of a linear interpolation of $V-K_{\rm s}$ and $G-K_{\rm s}$ vs.\ spectral type using online tables provided by E. Mamajek.
For stars with spectral types available in SIMBAD, we find agreement with the color-based interpolation within $\sim 2$ spectral subtypes. As expected, the vast majority of the stars have spectral types between mid-K and early-M, a result of the combination of the range of $G-K_{\rm s}$ over which they were selected and {\it Gaia} DR1 magnitude limits \citep[see][]{2017a_Kastner}.

 \subsection{Age diagnostics}

\subsubsection{UV excesses} 

\begin{figure}[h!]
\centering
\includegraphics[width=0.48\textwidth,angle=0]{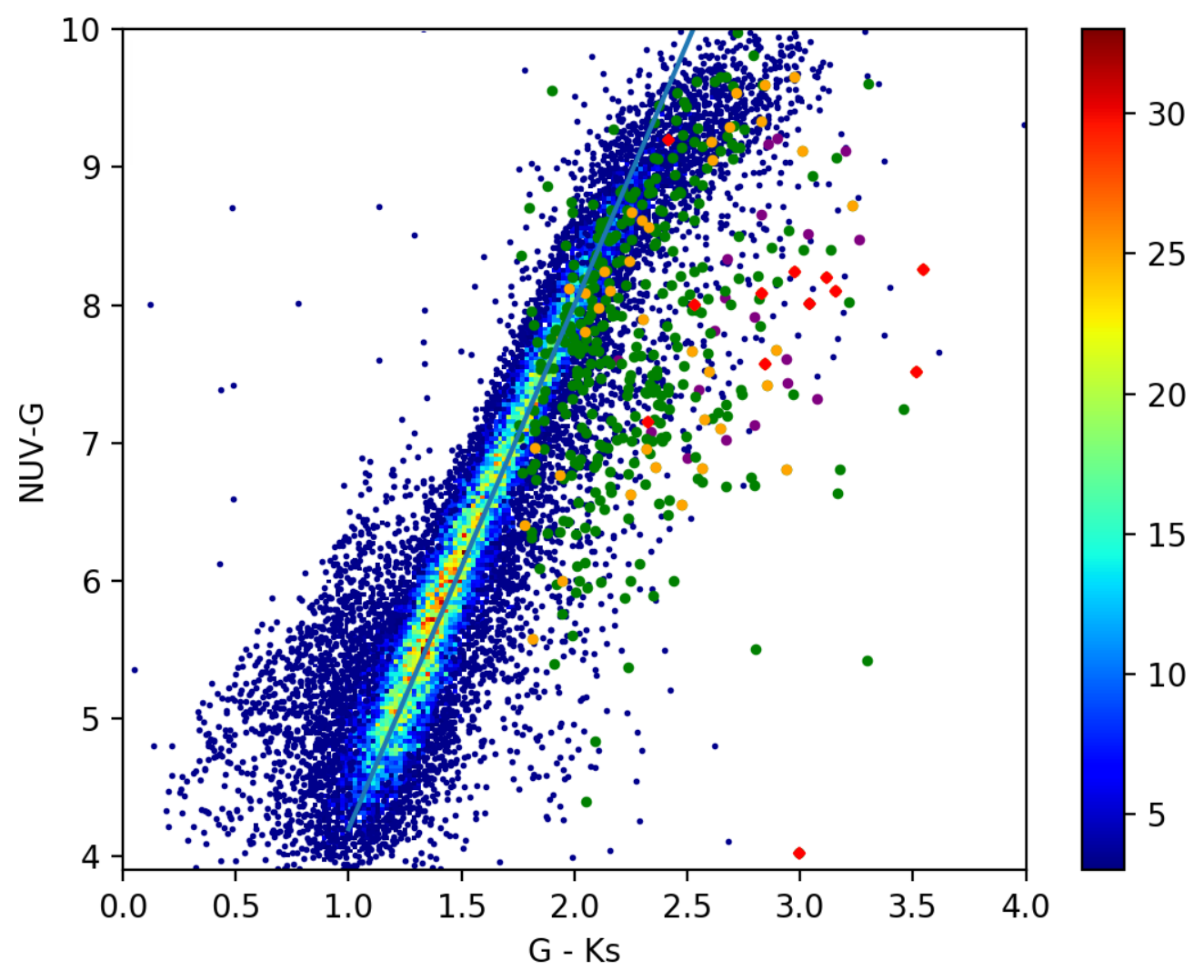}
\caption{
$NUV - G$ vs.\ $G-K$ color-color diagram for all {\it Galex}-selected stars in {\it Gaia} DR1, highlighting the positions of the 376 candidate stars (green symbols; yellow symbols for stars within 50 pc) as well as the sample of (19) stars from \citet[][purple symbols]{2017a_Kastner} and previously identified nearby, young stars (red symbols). The green line indicates the locus of main sequence stars; the displacement of a star to the right of this line is indicative of the presence and strength of its NUV excess.
\label{fig:NUVDiff}}
\end{figure}

In \citet{2017a_Kastner} it was demonstrated that UV-selected nearby young stars generally appear below the locus of main sequence stars in a $NUV-G$ vs $G-K_{\rm s}$ color-color diagram, due to their enhanced levels of chromospheric activity and (hence) near-UV excesses. Figure~\ref{fig:NUVDiff} confirms that the larger sample considered here adheres to this trend; the majority of the selected stars indeed appear to lie below the main sequence locus. 

\begin{figure}[h!]
\centering
\includegraphics[width=0.48\textwidth,angle=0]{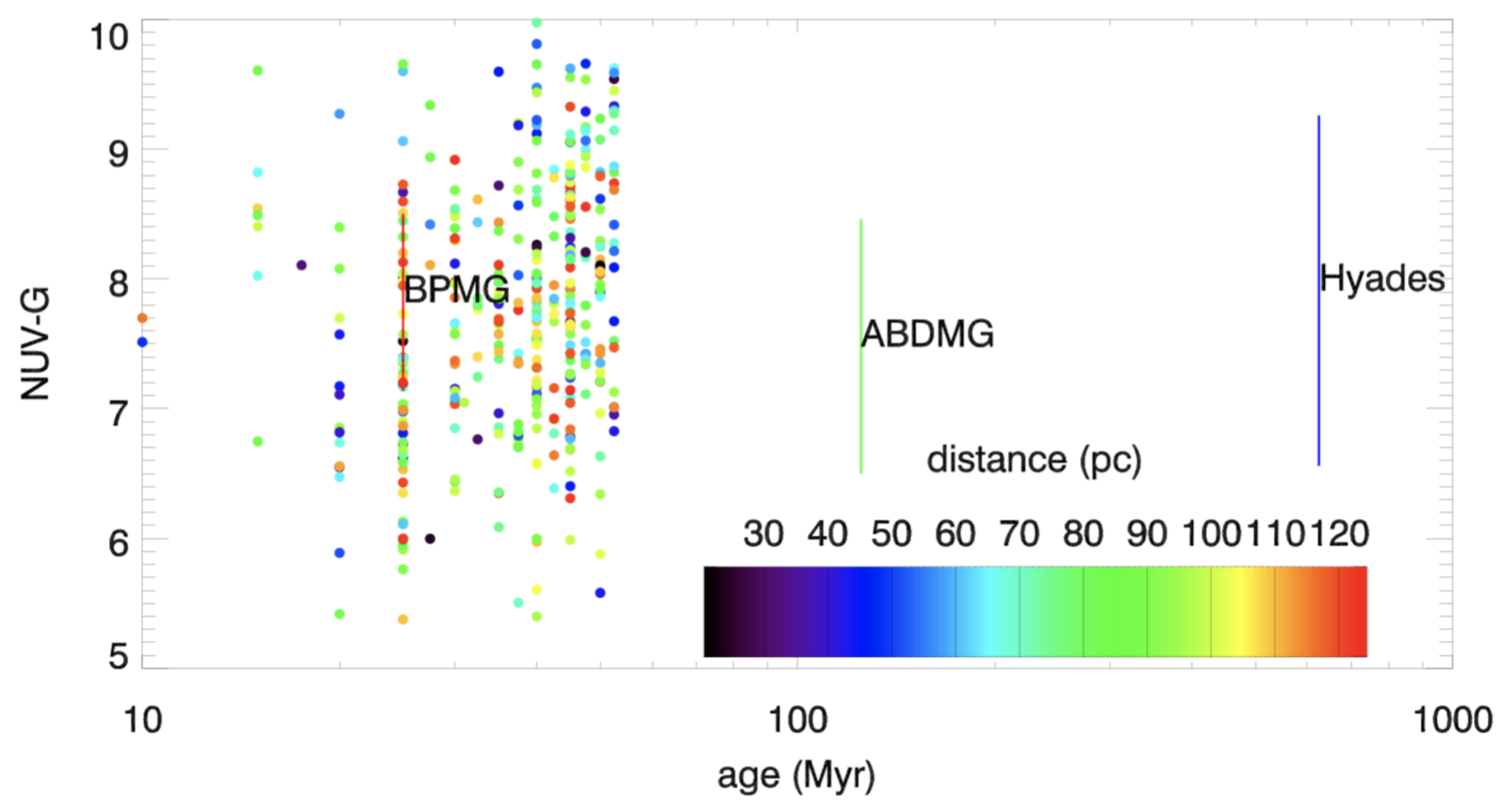}
\caption{
$NUV-G$ vs.\ isochronal age for the 376 candidate stars, with symbol color coding indicating distances to individual stars. The vertical line segments indicate the means and standard deviations of $NUV-G$ for the $\beta$ Pic Moving Group (BPMG), AB Dor Moving Group (ABDMG) and Hyades.
\label{fig:NUVvsAge}}
\end{figure}

In Fig.~\ref{fig:NUVvsAge}, we plot $NUV-G$ vs.\ isochronal age (and distance) for the 376 candidate stars. For reference, we overplot the means and standard deviations for K stars in the $\beta$ Pic Moving Group \citep[age $21-26\,$Myr;][]{2014a_Binks,2014b_Malo}, AB Dor Moving Group \citep[age $\sim 150\,$Myr;][]{2015a_Bell} and Hyades \citep[age $650 \pm 70\,$Myr;][]{2018a_Martin}. 
All three groups lie within $1\sigma$ of one another in $NUV-G$. Furthermore, the mean and standard deviation in $NUV-G$ for 217 K-type field-stars in the Gliese-Jahreiss catalog \citep{1991a_Gliese} is $8.40 \pm 1.07$, only slightly larger than (within $1\sigma$ of) the presumably younger NYMG samples. These statistics, along with Fig.~\ref{fig:NUVvsAge}, suggest that $NUV-G$ (or, by extension, UV excess) is of limited utility in isolating young stars from the field population.

\subsubsection{X-ray emission} 

Strength of X-ray emission (due to coronal activity) is another indicator of stellar youth;
it has long been recognized that pre-main sequence stars generally have measured values of $\log{L_{\rm X}/L_{\rm bol}}$ in the range $-4.0$ to $-3.0$ \citep[e.g.,][Binks \& Kastner 2019, in prep; and references therein]{1997a_Kastner,2016a_Kastner}. Of the 376 candidate young stars, only 103 (27.6\%) have ROSAT X-ray count rates listed in either the 1RXP, 2RXS or 2RXP (ROSAT All-Sky Survey; RASS) catalogs \citep{1999a_Voges,2000a_Voges,2016a_Boller}. While the ROSAT non-detection of the majority of candidate stars more distant than 50\,pc may partially reflect the RASS sensitivity limits \citep{2013a_Rodriguez}, the low RASS detection rate of the candidate stars within 50\,pc may provide contrary evidence for youth (see below). 

\begin{figure}[h!]
\centering
\includegraphics[width=0.48\textwidth,angle=0]{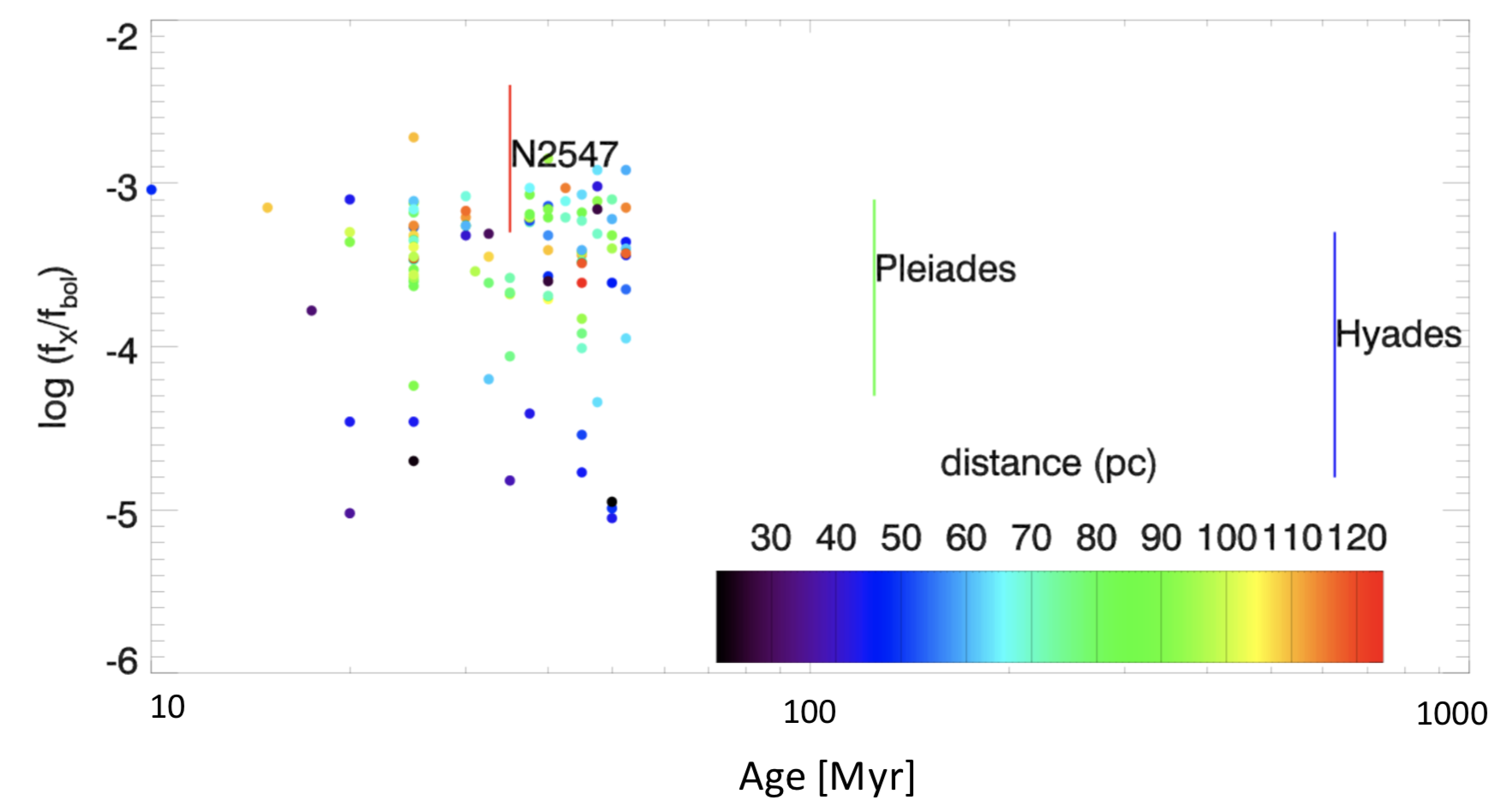}
\caption{
As in Fig.~\ref{fig:NUVvsAge}, but here we display $\log{f_{\rm X}/f_{\rm bol}}$ ($=\log{L_{\rm X}/L_{\rm bol}}$) vs.\ isochronal age and distance for the 103 candidate stars with RASS X-ray detections. The vertical line segments indicate the means and standard deviations of three young clusters (N2547 = NGC 2547).
\label{fig:XraysVsAge}}
\end{figure}

ROSAT count rates were converted to $f_{\rm X}$ as described in \citet{2016a_Kastner}, and bolometric fluxes ($f_{\rm bol}$) were estimated from the stars' spectral types and $J$ magnitudes using bolometric corrections listed in \citet{2013a_Pecaut}. The resulting plot of $\log{f_{\rm X}/f_{\rm bol}}$ ($=\log{L_{\rm X}/L_{\rm bol}}$) vs.\ age (and distance) for the 103 candidate stars with ROSAT X-ray detections is presented in Fig.~\ref{fig:XraysVsAge}, overlaid with the means and standard deviations of $\log{L_{\rm X}/L_{\rm bol}}$ for K stars in three nearby young star clusters (NGC 2547, Pleiades, and Hyades), to illustrate the decline of $\log{L_{\rm X}/L_{\rm bol}}$ over the age range 30 Myr to 650 Myr \citep[see also][]{1997a_Kastner}.  Comparison of Figs.~\ref{fig:NUVvsAge} and \ref{fig:XraysVsAge} provides tentative evidence that NUV emission remains elevated for K stars even after X-ray emission begins to decline, and that both the X-ray and UV distributions may broaden after K stars arrive on the main sequence. These results are consistent with those of \citet{2013a_Stelzer}, who found a similar relationship for UV and X-ray emission for M stars in NYMGs and in the field. 

Notably, the majority of our 103 candidate stars that were detected in X-rays by the RASS lie between the Pleiades and Hyades in $L_{\rm x}/L_{\rm bol}$, and many of the stars within $\sim$70 pc lie at or below the $L_{\rm x}/L_{\rm bol}$ level of the Hyades. Furthermore, of the 36 candidates within 50 pc, only 13 (36\%) were detected in the RASS. Hence, on the basis of X-ray emission characteristics, we would conclude that the majority of our 376 candidate young stars are very likely much older than their Gaia-based isochronal ages.

\subsubsection{H$\alpha$ emission, Li absorption, and IR excesses} 

The presence and strength of H$\alpha$ emission and equivalent width (EW) of Li absorption are well-established indicators of stellar youth \citep[e.g.,][and references therein]{2004a_Zuckerman}. There are H$\alpha$ measurements available in the literature for 47 candidates, of which only 6 display H$\alpha$ in emission. Thirty-six stars were found to have at least one Li EW measurement in the literature. This subsample is heavily biased toward stars previously identified as candidate NYMG members (see \S 3.3.2) and, indeed --- in contrast to the relatively small percentage of H$\alpha$ emitters --- 20 of these 36 stars have Li EWs that imply ages $<$150 Myr.

Evidence of a debris disk, in the form of an IR excess, is also an indicator of youth. Various studies have established that the general K/M field star population has a WISE $W1-W4$ color centered around $W1-W4=0$ with a dispersion of $\sim 0.3$, such that stars with $W1-W4 > 1.0$ are candidate debris disk hosts \citep[e.g.,][]{2017a_Binks}. There are two objects in our sample that satisfy this criterion: V1317 Tau (= 2MASS J$04234759+2940381$), previously identified as a weak-lined T Tauri star associated with the Taurus cloud \citep[][]{1996a_Wichmann}, and the infamous TW Hya (= 2MASS J$11015191-3442170$; see Table 1), with $W1-W4 =1.284$ and $W1-W4 = 5.585$, respectively.

\subsection{Kinematics}

\begin{figure*}[h!]
\centering
\includegraphics[width=0.33\textwidth,angle=0]{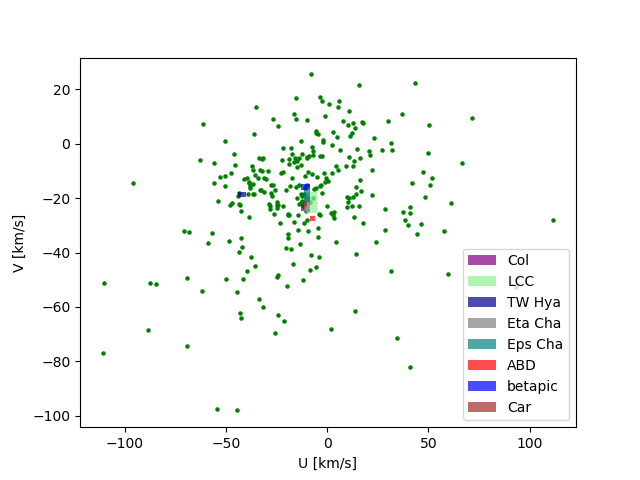}
\includegraphics[width=0.33\textwidth,angle=0]{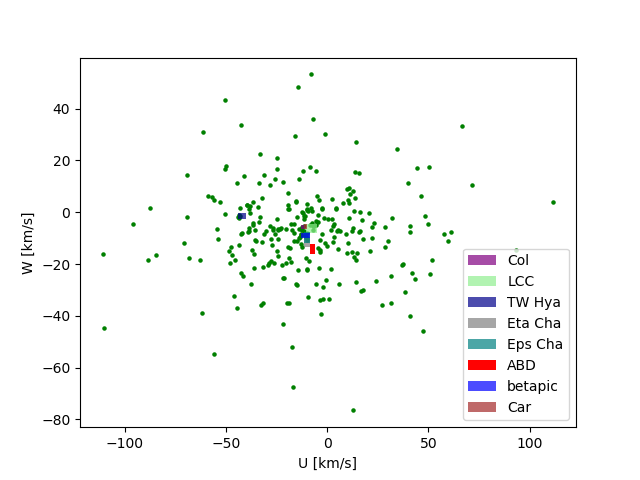}
\includegraphics[width=0.33\textwidth,angle=0]{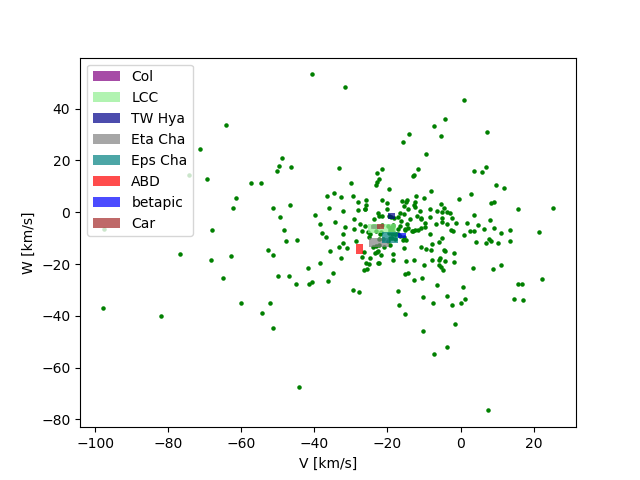}
\includegraphics[width=0.33\textwidth,angle=0]{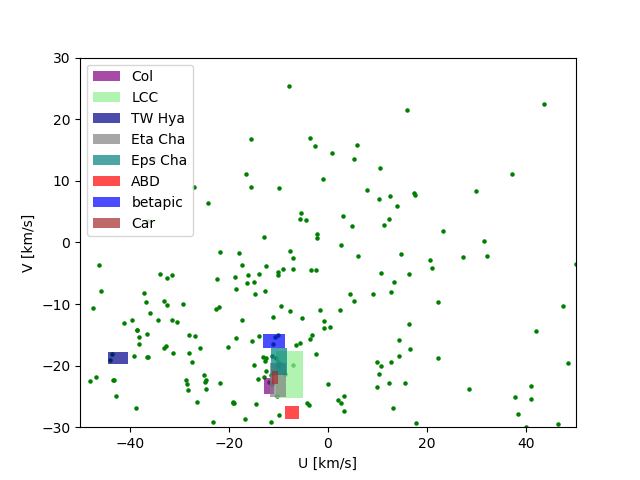}
\includegraphics[width=0.33\textwidth,angle=0]{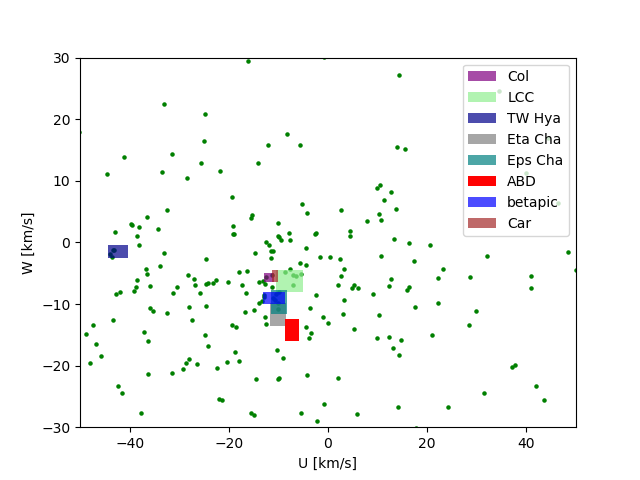}
\includegraphics[width=0.33\textwidth,angle=0]{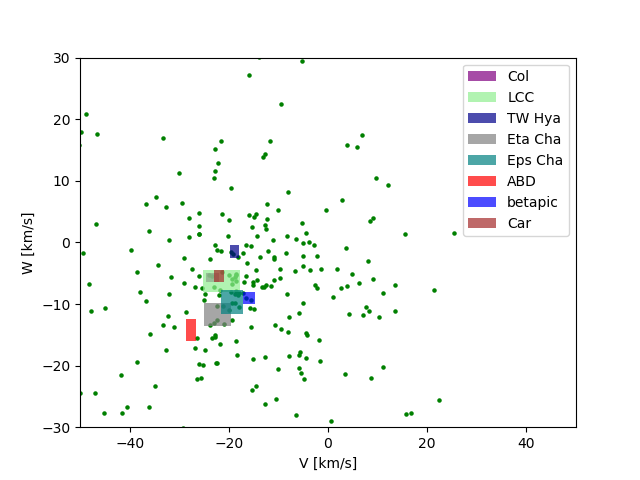}
\caption{$UVW$ space velocities of the nearby young star candidates selected as described in \S2, with centroid positions and extents of 8 well-established NYMGs \citep{2018a_Gagne} indicated. The axis ranges of the top row of panels are set  so as to display all 376 candidates, while the bottom panels provide zoomed-in views centered on $U=V=W=0$ and encompassing the NYMG centroids.}
\label{fig:UVWs}
\end{figure*}

\subsubsection{Space velocity distribution}

Galactic space velocities ($UVW$) and their errors were calculated from each star's position, proper motion, radial velocity (RV), and parallax (and their associated errors) following the prescription in \citet{1987a_Johnson}. All 376 {\it Gaia} DR1-selected objects in our candidate young star sample have position, proper motion, and parallax data available in the {\it Gaia} DR2 catalog. The median errors are $0.04095\,{\rm mas}$, $0.0805\,{\rm mas\,yr}^{-1}$ and $0.05230\,{\rm mas}$, respectively, where the positional and proper motion errors are the means of the RA and declination components. Distances are acquired from the Bayesian inference technique described in \citet{2018a_Bailer-Jones}. The RVs were sourced from the literature for the 285 stars with available measurements, using the VizieR online database. 

The resulting $UVW$ space velocities are displayed in Fig.~\ref{fig:UVWs}, overlaid with the centroid positions and approximate $UVW$ extents of 8 well-established NYMGs with ages up to $\sim$150 Myr \citep[selected from][]{2018a_Gagne}. It is immediately apparent that the vast majority of the 376 candidate nearby young stars have space velocities that place them far outside the individual and collective $UVW$ domains of these NYMGs. Thus, as in the case of the X-ray emission strengths of our candidates (\S 3.2.2), their kinematics strongly suggest that the majority of these candidates are older than 150 Myr. This condundrum is further explored in \S 4.

\subsubsection{NYMG membership}

To establish whether there exists a subset of our 376 candidate stars that are possible or established NYMG members, we applied a suite of kinematic membership tests: (1) a $\chi^{2}_{\rm MG}$ test on $UVW$ and associated errors --- with $\chi^{2}_{\rm MG}$ as defined in \cite{2012a_Shkolnik} and \cite{2015a_Binks} --- wherein we require $\chi^{2}_{\rm MG} < 3.78$, corresponding to 95\% probability of membership; (2) a kinematic distance test, in which the measured distance and the distance expected were the object a member of a given NYMG \citep[with NYMG distances defined via galactic position data provided in][]{2018a_Gagne}, $\Delta{\rm D}$, must agree to within 10\,pc; and (3) an RV test, in which the difference between the measured RV and the RV expected were the object a member of a given NYMG, $\Delta{\rm RV}$, must be less than $5\,{\rm km\,s}^{-1}$.
These criteria are applicable for all objects in the candidate young star list, regardless of whether there is an RV measurement or not; for stars lacking an RV, $\chi^{2}_{\rm MG}$ was calculated over a range of RVs to determine whether any plausible value provides a potential match to a known NYMG. In addition to these kinematic criteria, one must ensure that the isochronal age of the star is consistent with the age range of the NYMG membership that is implicated kinematically, and --- in cases where the implicated NYMG has a compact footprint in RA and declination \citep[e.g.,][their Fig.~1]{2018a_Gagne} --- that the candidate's position on the sky does not obviously disqualify it from membership.

Table~\ref{tbl:NYMGcandidates} summarizes the key results of applying the foregoing criteria to the 376 candidate stars. Among the 285 candidates with RV measurements, we find 19 stars satisfy the $\chi^{2}_{\rm MG}$ test, and 14 of these pass all further kinematic and sky position tests. Six of these 14 stars (2MASS J 04392545$+$3332446, 09503676$-$2933278, 10374741$-$0623225, 11212188$-$4736028, 13065028$-$4609561, and 22424884$+$1330532) have not previously been identified as members of a NYMG; another six have NYMG membership membership assignments in the literature that agree with our assignment of host NYMG; and our kinematic NYMG membership assignments for another two stars conflict with previous literature assignments. 
In addition, we find two stars that lack RV measurements but would be viable AB Dor Moving Group candidates if their RVs lie within a narrow range; these ranges hence serve as RV predictions for membership. Finally, we identify another half-dozen stars among our 376 candidates that have previously been assigned NYMG memberships in the literature, but that either fail at least one of our kinematic tests, or cannot be confirmed as members via our methods.

\section{Discussion}

Our UV- and {\it Gaia}-DR1-based search for nearby ($D \le 120$ pc), young (age $< 100$ Myr) stars has yielded the potential identification of 6--8 new candidate members of NYMGs and the recovery of another dozen or so previously known NYMG members (\S 3.3.2; Table 1). While these identifications demonstrate the potential of the methods described here, the rather low yield --- roughly 20 known or new NYMG members among 376 candidates --- is surprising, given that our 376 candidates were selected on the basis of {\it Gaia}-based isochronal ages $\le$80 Myr (Fig.~\ref{fig:GaiaColMag}) as well as large UV fluxes (Fig.~\ref{fig:NUVDiff}). Similarly, Figs.~\ref{fig:XraysVsAge} and \ref{fig:UVWs} provide strong evidence that our sample of nearby young star candidates is in fact dominated by stars with ages $>$150 Myr, despite the fact that these stars lie high above the locus of MS stars in a {\it Gaia}/2MASS color-magnitude diagram (Fig.~\ref{fig:GaiaColMag}).  

These results have far-reaching implications for the use of {\it Gaia}-based isochronal ages to select candidate young stars for purposes of imaging giant exoplanets, as well as for models of the manifestation and evolution of stellar magnetic activity. 
Before further considering these implications, we mention three somewhat unlikely explanations for the non-young-star-like X-ray and kinematic properties of the majority of the candidates.\\ 
(1) {\it Contamination by first-ascent giant stars.} This explanation would appear to be at odds both with the CMD distribution of the candidates (Fig.~\ref{fig:GaiaColMag}) and with the frequency of UV excesses among the candidate stars (Fig.~\ref{fig:NUVDiff}). \\
(2) {\it Binaries with a narrow range of separations.} Binary stars with separations such that both stars are included in 2MASS photometry ($\sim 2''$ PSF), but only the primary is measured by {\it Gaia} at $G$ band, would shift an apparently single star upwards and to the right in Fig.~\ref{fig:GaiaColMag}.
However, such confusion should only apply to a highly specific subsample of binary stars with separations around $\sim$1$''$. \\
(3) {\it MS stars with white dwarf (WD) companions.} A subset of our candidates may be MS stars that have been spun up and/or inflated by accretion of mass lost by the asymptotic giant branch antecedent of a companion WD \citep{Jeffries1996}, such that the WD is in fact a contributor to (or dominates) the UV detected by {\it Galex}. But it seems highly improbable that such systems would dominate our sample. 

The surprising ``bifurcation'' of our UV- and isochronally-selected nearby young star candidates into X-ray-bright and X-ray-faint subsamples (Fig.~\ref{fig:XraysVsAge}) raises a set of particularly vexing questions. Namely: If the large (apparently dominant) fraction of X-ray-faint stars among our candidates are in fact not pre-MS stars, then why are they as ``overluminous'' (in terms of their bolometric luminosities) as the X-ray-bright stars? Are they ``imposters'' --- zero-age (or even older) main sequence (MS) stars that are ``puffed up'' via high levels of magnetic activity due to fast rotation rates, as appears to be the case for, e.g., the subset of overluminous late-type Pleiades stars \citep{SomersStassun2017}? 
If so, why do so many of our magnetically active, presumably radially inflated young MS field stars have such weak coronae relative to ``normal'' young stars, despite their apparently similar levels of chromospheric UV excess? Could some or all of these stars be rotationally inflated yet X-ray faint due to centrifugally induced ``coronal stripping'' \citep{Jardine2004}? 

Addressing these questions will require a dedicated observing campaign targeting our candidate stars in the optical through UV to X-ray regimes, so as to access diagnostics of chromospheric and coronal activity and relate these properties to stellar age indicators and rotation rates. To that end, we have been obtaining moderate- to high-resolution optical spectra of a representative sample of the candidates (Chalifour et al.\ 2019, in prep.). Ultraviolet spectroscopy with HST, as well as Chandra and XMM X-ray spectroscopy, represent the additional, essential puzzle pieces necessary to understand the natures of the class of isochronally young and UV-bright yet X-ray faint and kinematically ``old'' stars uncovered by this work. Regardless, the preliminary work presented here, like that of  \citet[][]{WrightMamajek2018}, hints at the power of {\it Gaia}  astrometric and photometric data for purposes of isolating the population of young MS stars that originated in young moving groups and have relatively recently mixed into the field star population.

\subsection*{Acknowledgments}

This research was supported by NASA Astrophysics Data Analysis Program (ADAP) grant NNX12AH37G to RIT and UCLA and NASA ADAP grant NNX09AC96G to RIT, and by a National Science Foundation Research Experience for Undergraduates program grant to RIT's Center for Imaging Science that supported M. Chalifour's summer 2017 RIT residency. 

\begin{center}
\begin{table*}[htb]
\footnotesize
\caption{\sc Candidate NYMG Members}
\label{tbl:NYMGcandidates}
\medskip
\begin{tabular}{cccccccccc}
\hline
\hline
Name & Isochronal Age & Distance & NYMG$^a$ & $\chi^{2}$ & $\Delta{\rm RV}$ & $\Delta{\rm D}$ & ${\rm NYMG_{\rm lit}}$$^a$ & ref(s)$^b$ & BANYAN$^c$ \\
(2MASS J-) & (Myr) & (pc) & & & (${\rm km\,s}^{-1}$) & (pc) \\
\hline
\multicolumn{10}{c}{Candidates with a measured RV and with $\chi^{2} < 3.78$} \\
\hline
02303239$-$4342232 & $20-60$	& $52.579_{-0.663}^{+0.680}$	& (TWA)$^d$	& $0.802$	& ...	& ...	& COL	& abc	&           \\
02414683$-$5259523 & $<10-30$	& $43.756 \pm 0.072$		& THA	& $1.629$	& $-1.987$	& $1.117$	& THA	& abcd	& THA(99.9) \\
04090973$+$2901306 & $<10-40$	& $109.965_{-0.503}^{+0.508}$	& (TWA)$^d$	& $0.809$	& ...	& ...	&	&	& TAU(13.7) \\
04392545$+$3332446 & $20-60$	& $89.859_{-0.365}^{+0.367}$	& THO	& $1.894$	& $-2.167$	& $2.528$	&	&	&           \\
05004714$-$5715255 & $20-60$	& $26.880 \pm 0.020$		& BPM	& $0.461$	& $+0.552$	& $2.034$	& BPM	& abcd	& BPM(99.9) \\
05214684$+$2400444 & $10-50$	& $88.044_{-0.319}^{+0.321}$	& THO	& $3.360$	& $-1.511$	& $25.403$	&	&	&           \\
09503676$-$2933278 & $15-70$	& $120.650_{-0.527}^{+0.531}$	& BPM	& $1.017$	& $-1.533$	& $8.202$	&	&	&           \\							
                   &            &                               & EPS	& $2.567$	& $-4.168$	& $14.123$	&	&	&           \\
10182870$-$3150029 & $25-70$	& $65.520 \pm 0.139$		& TWA	& $2.082$	& $+3.543$	& $4.676$	& TWA	& abd	&           \\
                   &            &                               & COL	& $2.683$	& $+0.409$	& $2.995$	&	&	&           \\
10374741$-$0623225 & $15-60$	& $37.824 \pm 0.059$		& OCT	& $1.672$	& $-1.482$	& $5.663$	&	&	&           \\
11015191$-$3442170$^e$ & $<10-20$	& $59.982_{-0.150}^{+0.151}$	& TWA	& $0.479$	& $-0.326$	& $1.741$	& TWA	& abcd	& TWA(99.9) \\
11212188$-$4736028 & $<10-40$	& $70.579 \pm 0.160$		& BPM	& $3.609$	& $+3.037$	& $17.407$	&	&	&           \\
                   &            &                               & ABD	& $2.872$	& $-7.798$	& $2.456$	&	&	&           \\
                   &            &                               & OCT	& $3.188$	& $+15.910$	& $20.829$	&	&	&           \\
                   &            &                               & EPS	& $1.020$	& $+0.619$	& $7.648$	&	&	&           \\
11315526$-$3436272 & $<10$	& $49.309_{-0.143}^{+0.144}$	& TWA	& $0.021$	& $+0.099$	& $0.418$	& TWA	& abcd	& TWA(99.9) \\
11594226$-$7601260 & $<10-30$	& $99.460_{-0.230}^{+0.232}$	& EPS	& $0.047$	& $-0.102$	& $0.094$	& CAN	& d	& EPS(99.9) \\
13065028$-$4609561 & $20-60$	& $98.621_{-0.499}^{+0.504}$	& TWA	& $0.468$	& $-3.802$	& $5.261$	&	&	& LCC(28.7) \\
                   &            &                               & BPM	& $1.584$	& $-1.934$	& $3.606$	&	&	& UCL(63.3) \\
                   &            &                               & CAR	& $3.209$	& $-7.979$	& $6.679$	&	&	&           \\
14015830$+$1925296 & $10-40$	& $64.077_{-0.132}^{+0.133}$	& (OCT)$^f$	& $2.294$	& $+6.030$	& $55.606$	&	&	&           \\
14590325$-$2406318 & $20-80$	& $113.447_{-0.452}^{+0.456}$	& TWA	& $2.737$	& $+4.294$	& $24.248$	&	&	& UCL(92.0) \\
                   &            &                               & BPM	& $2.541$	& $+5.072$	& $7.806$	&	&	&           \\
21351099$+$3402313 & $15-70$	& $75.599_{-0.182}^{+0.183}$	& (OCT)$^f$	& $0.972$	& ...	& ...	&	&	&           \\
22424884$+$1330532 & $<10-20$	& $67.620_{-0.383}^{+0.387}$	& COL	& $3.345$	& $+2.006$	& $7.162$	&	&	&           \\
23323085$-$1215513 & $25-70$	& $27.352 \pm 0.044$		& BPM	& $0.499$	& $+0.660$	& $0.020$	& BPM	& abc	& BPM(99.9) \\
\hline
\multicolumn{10}{c}{Candidates without a measured RV and with $\chi^{2} < 3.78$} \\
\hline
02552739$+$5020228 & $15-70$	& $117.742_{-0.558}^{+0.563}$	& ABD	& \multicolumn{2}{c}{ ($-13.0,-10.0$)$^g$}			& $0.243$	&	&	&           \\
04310860$+$6445082 & $20-80$	& $43.807_{-0.164}^{+0.166}$	& ABD	& \multicolumn{2}{c}{ ($+10.0,+12.5$)$^g$}			& $0.502$	&	&	&           \\
\hline
\multicolumn{10}{c}{Candidates assigned as NYMG members in literature but rejected/omitted by our criteria} \\
\hline
00565546$-$5152319 & $<10-50$	& $37.036_{-0.275}^{+0.279}$	& ARG	& $>100$	& $-3.683$	& $23.599$	& ARG	& abc	&           \\
03315564$-$4359135 & $25-80$	& $45.216 \pm 0.071$		& THA	& $9.169$	& $+1.196$	& $1.919$	& THA	& abc	&           \\
04480066$-$5041255 & $20-60$	& $59.499_{-0.269}^{+ 0.272}$	& THA	& $33.464$	& $+6.042$	& $43.723$	& THA	& abc	&           \\
04524951$-$1955016 & $10-60$	& $61.576_{-0.097}^{+ 0.098}$	& THA	& $87.503$	& $+44.000$	& $14.899$	& THA	& c	&           \\
 12151838$-$0237283 & $25-80$	& $51.515_{-0.217}^{+ 0.219}$	& ABD	& $3.833$	& $+1.627$	& $3.955$	& ABD	& a	&           \\
23002791$-$2618431 & $10-25$	& $31.830_{-0.049}^{+ 0.050}$	& ABD	& $5.253$	& $+3.471$	& $1.450$	& ABD	& abcd	& ABD(98.8) \\
\hline
\end{tabular}
\smallskip

NOTES: \\
a) ABD =  AB Doradus MG, ARG = Argus, BPM = Beta Pictoris MG, 
COL = Columba MG, EPS = epsilon Cha 
LCC = Lower Centaurus Crux,  OCT = Octans-Near MG, TAU = Taurus, THA = Tucana-Horologium Association,
THO = 32 Ori MG, TWA = TW Hydrae Assoc., UCL = Upper Centaurus Crux. \\
b) References: a = \citet{2015a_Bell}; b = \citet{2013a_Malo}; c = \citet{2008a_Torres}; d = \citet{2004a_Zuckerman}. \\
c) BANYAN $\Sigma$ probability of membership \citep{2018a_Gagne}. \\
d) Kinematic $\chi^{2}$ match to TWA, but sky position incompatible with TWA membership. The star 04090973$+$2901306 is most likely associated with the Taurus cloud. \\
e) TW Hya. \\
f) Kinematic $\chi^{2}$ match to OCT, but sky position incompatible with OCT membership. \\
g) Viable RV range for membership.
\end{table*}
\end{center}


\end{document}